# Synchrotron X-ray Studies of Superlattice Ordering in $Pb(Mg_{1/3}Nb_{2/3})O_3$ Single Crystals Doped with $PbTiO_3$

Andrei Tkachuk[*], Paul Zschack[*], Eugene Colla[†] and Haydn Chen[*]

[*]*Department of Materials Science and Engineering and Frederick Seitz Materials Laboratory, University of Illinois, Urbana, Illinois 61801, USA*

[†]*Department of Physics, University of Illinois, Urbana, Illinois 61801, USA*

**Abstract.** The temperature dependence of the superlattice reflections: a) F spots and b) $\alpha$ spots in a lead magnesium niobate (PMN) single crystals containing 0% and 6% of $PbTiO_3$ (PT) has been studied using synchrotron x-ray scattering techniques. (No superlattice reflections were found in PMN doped with 32% PT). Analysis of the temperature dependence of the $\alpha$ spots suggests the existence of the correlated anti-parallel atomic displacements that form nanoregions different from the chemical nanodomains. While the correlation length is temperature independent, the magnitude of these displacements increases on cooling below the freezing temperature $T_f$. Intensities of the $\alpha$ spots above this temperature become indistinguishable from the background. Our results show that value of $T_f$ for each composition is very close to the one obtained from a Vogel-Fulcher fit to the frequency dependence of the dielectric constant maximum $T_m$. The relation of these correlated anti-ferrodistortive fluctuations to polar ferroelectric nanodomains and relaxor behavior needs further study.

## INTRODUCTION

High dielectric constant of the relaxor ferroelectrics, which can exceed 50,000, is an attractive feature for applications in the capacitor industry. Relaxor properties can be tuned by different atomic substitutions. This was shown to increase material's electrostriction and piezoelectric coefficients, thereby expanding the applications to the area of very effective actuators and nanopositioners [1]. Among other interesting characteristics of the ferroelectric relaxors, the following defining properties necessitate better understanding: a) broad dielectric constant peak over a wide temperature range, b) strong dispersion of the complex susceptibility as a function of frequency of the applied electric field. These features which distinguish the relaxors from the "normal ferroelectrics" are presently under extensive investigation. As a result, several models have been proposed in publications to explain the origin of the relaxor behavior [2-5]

$Pb(Mg_{1/3}Nb_{2/3})O_3$ (PMN) can be considered as a model ferroelectric relaxor. The temperature dependence of the dielectric constant exhibits a broad peak near $T_m$=265 K (AC f=1 kHz). Its position and magnitude strongly depend on the frequency of the applied electric field [3]. Substitution of Ti on the B sites (Mg/Nb) of the $ABO_3$ perovskite structure of the PMN, was shown to change the dielectric properties of the material. In general, this type of doping reduces the frequency dispersion of the dielectric constant and moves the position of the peak's maximum ($T_m$) to higher temperatures. Relaxor properties disappear for concentrations of Ti apporaching 34% which marks the morphotropic phase boundary on PMN-PT diagram.

Present understanding of relaxors assumes the existence of small regions of the ferroelectric polar clusters, which may interact with each other by means of dipolar interactions [3]. The size of these polar regions is on the nanometer scale. They were postulated to exist at temperatures much higher than $T_m$ and grow in size during cooling [3]. Polar regions are imbedded into a chemically disordered matrix of Mg and Nb atoms. At the same time Nb and Mg are also ordered on the nano-scale, forming chemically ordered nanodomains. The random layer model, proposed by Davis et al [6], suggests the most probable type of chemical ordering. Chemical disorder is believed to prevent polar nanodomains from growing into regular micro-size ferroelectric domains [7,8]. At the sufficiently low freezing temperature ($T_f$), the polar clusters become randomly frozen in space without forming long-range ordered ferroelectric phase [2]. This scenario is very similar to the magnetic spin glasses where the structural disorder coupled with competing interactions accounts for observed glassy behavior [9]. However, phenomena related to disorder in the dielectrics are more complicated than in magnetic systems. Electric charge interactions for example, may cause large structural distortions, which do not take place in magnetic spin glasses. In magnetic systems the source of competing interactions is due to frustration when both ferromagnetic and anti-ferromagnetic interactions occur simultaneously. It is still an open question if there is a source of competing interactions in the Pb containing relaxors. From our measurements, there is evidence of correlated anti-parallel distortions forming small nanoclusters. These exist in pure PMN and PMN-6%PT on a short-range scale. The magnitude of these distortions grows on cooling below the $T_f$. Therefore, the distortions might be considered a source of intractions competing with the establishment of ferroelectric order. However, the aforementioned is still somewhat speculative.

Local disorder and local fluctuations (chemical and dipolar) in PMN and other related relaxors, was demonstrated to be intimately related to relaxor ferroelectric behavior [3,7,10-13]. Correlation in these fluctuations is responsible for the formation of very weak and broad superlattice reflections. They were studied by utilization of synchrotron x-ray scattering techniques. The focus of this study was to understand superlattice reflections as a function of temperature and limited chemical compositions currently available to us.

## EXPERIMENTAL

Single crystal of PMN was grown by the Chochralsky technique. PMN-6%PT and PMN-32%PT single crystals were grown by the melted flux method. Sizes of the single crystals were $3x5x1\ mm^3$ for pure PMN, $5x5x0.5\ mm^3$ for PMN-6%PT and $7x7x0.5\ mm^3$ for PMN-32%PT with its surface normal perpendicular to 001 crystallographic plane. All samples were thoroughly studied and characterized with dielectric measurements of the complex dielectric constant as a function of frequency and temperature. We studied the crystals with synchrotron x-ray radiation at beamline X-18A at the National Synchrotron Light Source (NSLS), Brookhaven National Laboratory and beamline 33-ID at the Advanced Photon Source (APS), Argonne National Laboratory. Crystals were studied at the NSLS with the incident x-ray energy tuned to 10 keV by two Si (111) single crystals and subsequently focussed by the cylindrical mirror. Measurements were performed in the reflective geometry with KEVEX solid state detector. At APS the x-ray energy was tuned by a double-crystal monochromator using Si 111 and focussed vertically by two Rh coated mirrors. Diffracted radiation was measured with an Oxford scintillation detector. Additional studies near Pb $L_{III}$ absorption edge (13.35 keV) and Nb K edge (18.99 keV) were performed at APS on pure PMN. In both NSLS and APS, harmonics in the incident x-ray beam were suppressed to the level that did not cause any observable data contamination. Low temperature measurements were performed in a closed cycle He compressed gas cryostat, which was mounted on the standard four-circle diffractometer. The sample height was regularly adjusted to compensate for the temperature contraction of the sample holder inside the cryostat.

# RESULTS AND DISCUSSION

The temperature $T_m$ does not mark any structural phase transition. Numerous electron, x-ray and neutron scattering experiments reported in the literature were unable to find any clear evidence of macroscopic phase transition in a wide 15-600 K temperature range in powder and single crystals. These results were obtained from studies of fundamental Bragg reflections. Bragg reflections are mostly sensitive to the average long-range structure. Superlattice short-range order peaks and diffuse scattering are more sensitive to the local structure. Two types of very weak superlattice reflections were found to exist in both pure PMN and PMN-PT for small concentrations of Ti: a) F peaks and b) α peaks [14]. The Cubic reciprocal unit cell is shown in Figure 1 where F and α spots are marked.

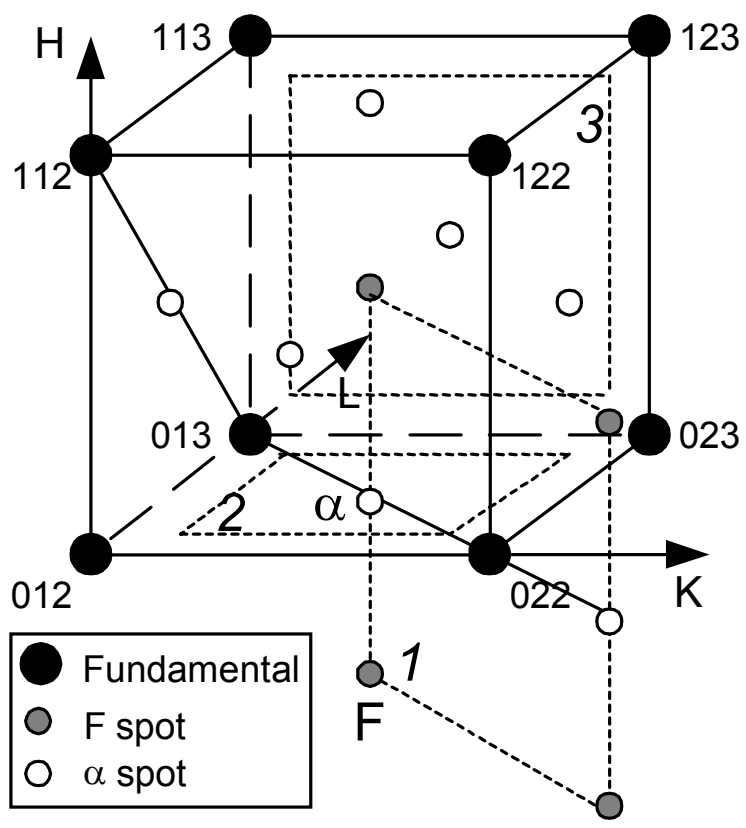

**FIGURE 1.** Reciprocal unit cell: corners-Bragg peaks; body centers- F spots and face centers-α spots. Reciprocal planes, discussed in the text, are labeled 1, 2 and 3. Dashed lines mark their perimeters.

Fundamental Bragg peaks, F spots and α spots are indicated by circles in Figure 1. Doubling of the unit cell in the real space leads to the half integer indexing of the superlattice reflections in the reciprocal space. Terminology for calling these peaks F and α spots is adapted from electron microscopy studies of the related ferroelectric relaxors [14]. Superlattice reflections were found initially in pure PMN and PMN-PT by transmission electron microscopy (TEM) techniques [14,15]. Later Vakhrushev et al. [16] found both F and α spots from the x-ray synchrotron measurements, but no temperature dependence of these reflections was reported. Zhang et al. [17] have studied the temperature dependence of the F spots with a rotating anode x-ray generator, but they were unable to observe any α spots. This is not surprising since α spots are weaker and more diffused than F spots so that the incident intensity from a conventional x-ray source might not be high enough to resolve them from the background. Synchrotron radiation is much brighter than conventional x-ray sources and therefore more useful to study these weak peaks. Gosula et al. [18] were the first to study the temperature dependence of F and α spots in pure PMN using synchrotron radiation. Work reported here extends the studies from pure PMN to PMN-6%PT and PMN-32%PT compositions.

In this work we have obtained the 3D intensity maps in sections of the reciprocal space. Figure 2 depicts the intensity distribution near the base of the 022 fundamental Bragg peak in the plane marked 1 in Figure 1. F and α spots are clearly seen in pure PMN and PMN-6%PT at 15 K. They are very weak in comparison to 022 Bragg peak, whose intensity is more than 8 orders of magnitude higher. We found the F peaks in PMN-6%PT to be weaker and broader than in pure PMN (see Figure 2). The width of the α spots for both compositions is identical, but the temperature dependence of the integrated intensity is different and will be discussed later in this section. No superlattice reflections of any kind were found in PMN-32%PT (T=15-300 K).

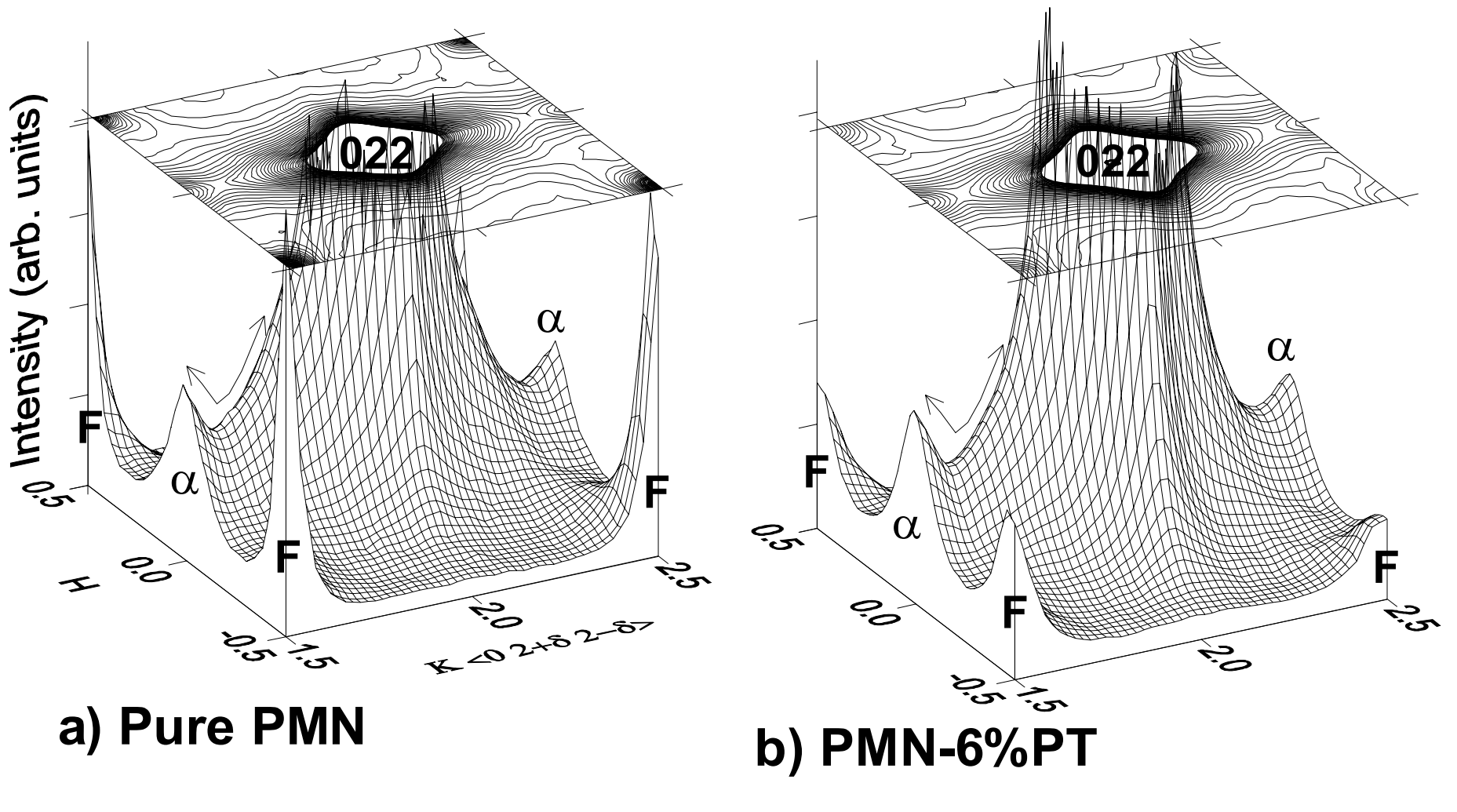

**FIGURE 2.** Section of reciprocal space marked 1 (Fig 1) in the vicinity of 022 Bragg peak. Figure indicates the presence of F and α spots for both PMN and PMN-6%PT at 15 K. The double-ended arrow indicates direction of the diffuse scattering ridge.

The main contribution to the F peaks is coming from the Nb/Mg chemical short-range ordering in [111] directions [14,15,17-19]. Additional contributions may come from the corresponding atomic displacements, as was proposed by Zhang et al. [17]. However, the effect is much smaller compared to that due to chemical ordering. It appears that $PbTiO_3$ doping destroys the Mg/Nb short range ordering. The estimated size of the chemically ordered regions in PMN-6%PT was obtained from the full width at half maximum (FWHM) using the well-known Scherrer equation for crystallite size determination. It was found on average to be 24 Å. This is similar to the size obtained from the electron diffraction dark field imaging in PMN doped with 7% Ti, where the estimated size was reported to be close to 30 Å [14]. The size of chemically ordered domains in pure PMN is about 50 Å. The intensity of the α peaks was found to be strongly temperature dependent in contrast to TEM studies of Hilton et al. [14]. This can be explained by the fact that additional intensity from the diffuse scattering ridge, like the one marked by arrow in Figure 2, contributes to the intensity of the α spots as a background. Therefore, any cross section of the ridge will appear as a peak. In Figure 1 these ridges are shown schematically by solid lines that stretch between the corners of the cube along their face diagonals. A different cross section of the same ridge in plane 2 (see Fig. 1) is shown in Figure 3. Figure 3a clearly shows coexistence of the α spot with a diffuse ridge at low temperature. It is clear that the α spot can be completely separated from the diffuse scattering only along the direction of the ridge. Linear scans across the ridge can be mistaken for a peak even if the actual α spot is no longer present (Figure 3b). This may be why earlier TEM measurements reported α spots to be temperature independent [14]. Figure 3b shows that the 1/2(035) α peak in PMN-6%PT cannot be resolved from the diffuse intensity ridge at 300 K. As discussed by H. You et al. [20], the origin of the diffuse scattering is attributed to the pure transverse soft phonon modes. Acoustic modes contribute to the intensity of the diffuse ridges near Bragg peaks with H+K+L=even, and optic modes contribute when H+K+L=odd [20].

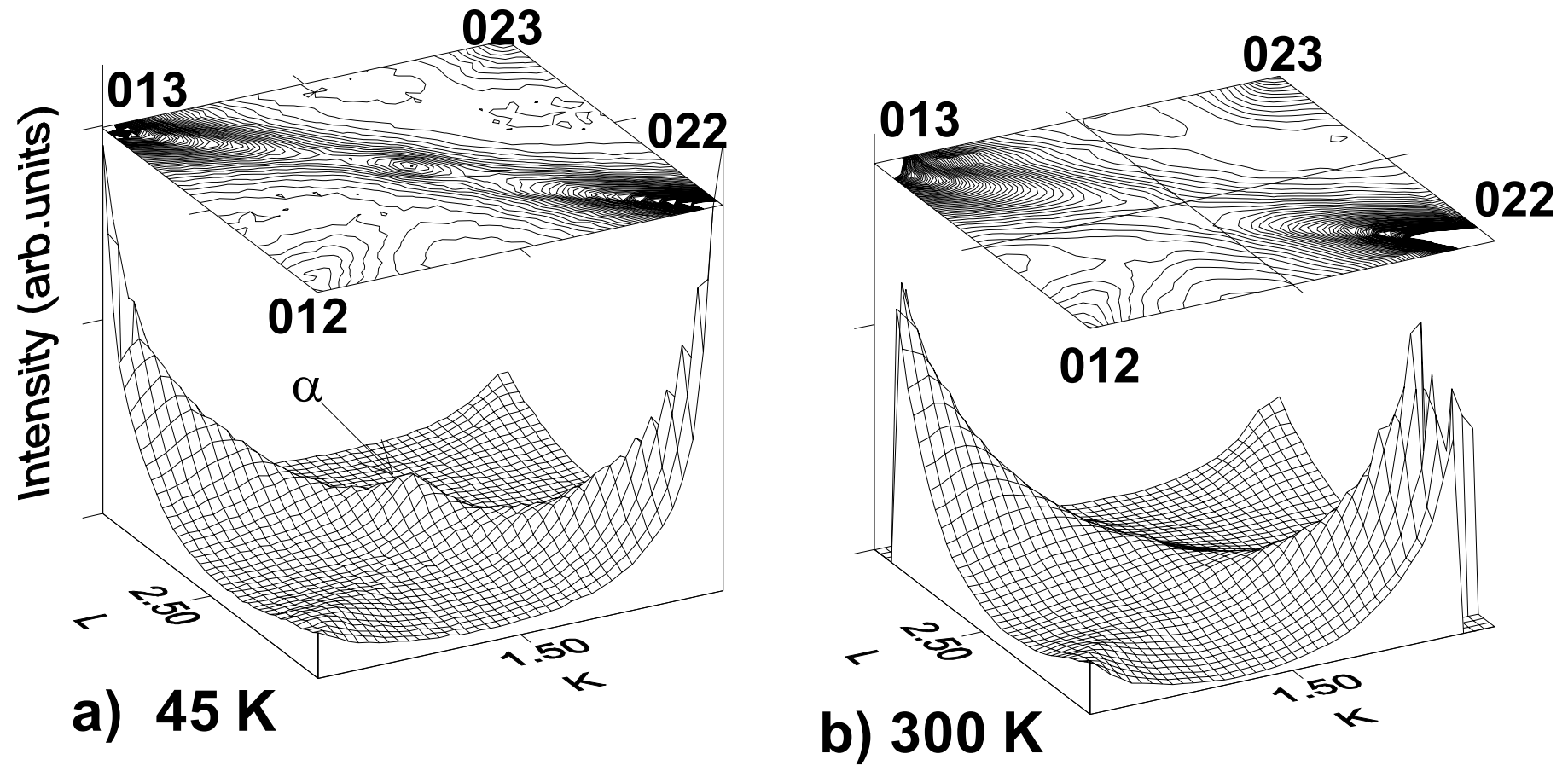

**FIGURE 3.** Section of reciprocal space marked 2 (Fig 1) in the vicinity of 022 Bragg peak. Figure indicates α spot at low temperature a) 45 K, which is indistinguishable from the diffuse intensity ridge at b) 300 K. Both 0% and 6% PT samples behave this way, but α spots disappear at different temperatures.

It has been shown that in the immediate vicinity of the Bragg peaks, the intensity of the diffuse scattering increases on cooling [16,20]. However, we found that closer to the center of the zone where α spots are located, ridges are not changing with temperature up to 800 K. Figure 4 depicts the temperature dependence of the integrated intensity of some α spots from the linear scan along the diffuse ridge. Different α spots of both pure PMN and PMN-6%PT are shown for comparison on the same plot normalized to their corresponding intensity at 15 K. To summarize, we found that the intensity of the α spot is decreasing on warming from 15 K, while the intensity of the ridge remains constant.

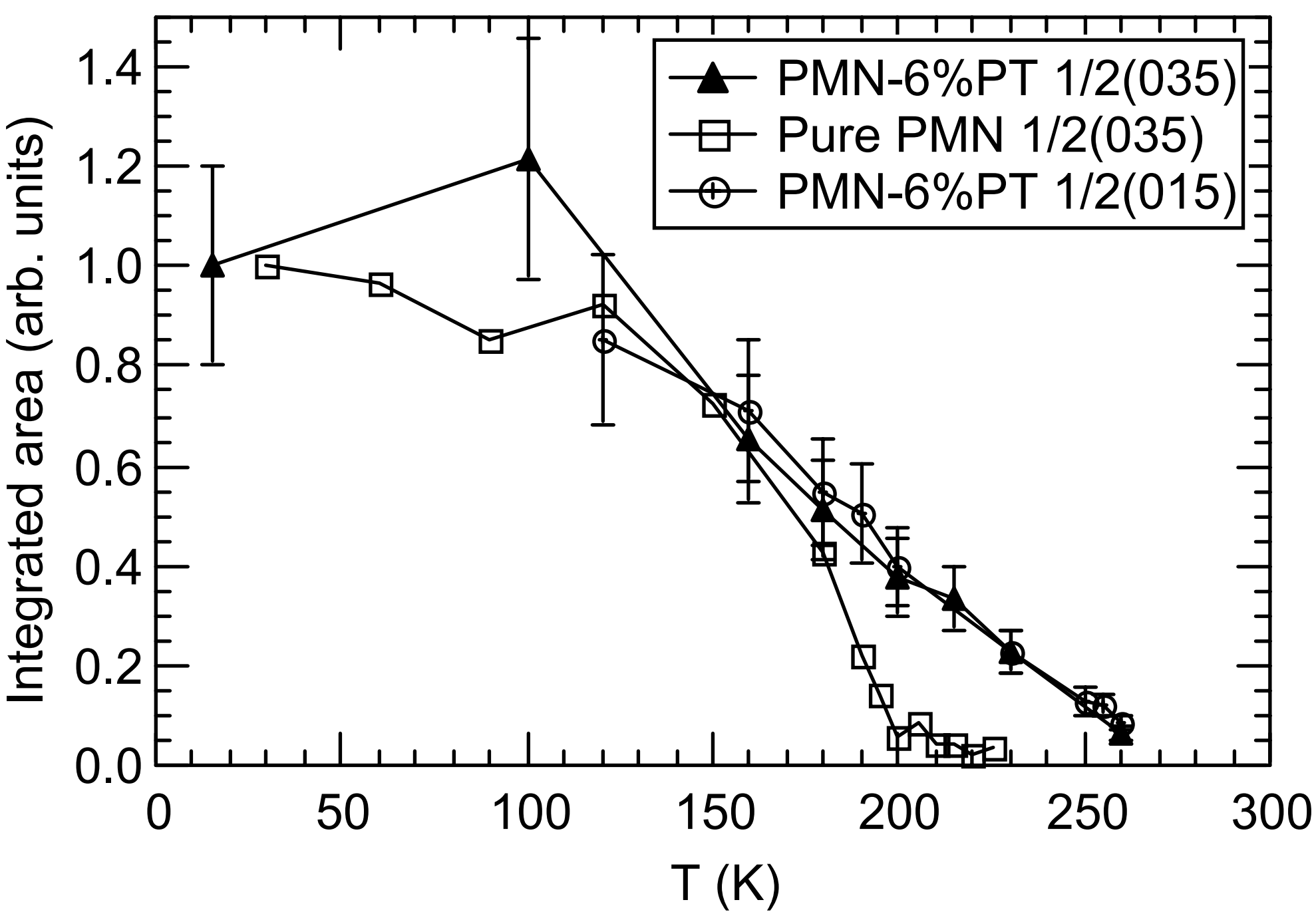

**FIGURE 4.** Temperature dependence of integrated intensity of α spots in PMN and PMN-6%PT.

Glazer [21] studied the origins of different superlattice reflections in perovskites. According to his proposed classification scheme, the origin of the α spots can be attributed to correlated in-phase oxygen octahedra tilts. The analysis of the structure factor has shown that α spots with two identical odd indices (e.g. 1/2(011), 1/2(033) etc.) cannot occur due to the oxygen tilting alone. In our experimental study we were able to observe such reflections. For example, 1/2(055) a spot is shown in Figure 5 This means that correlated displacements of atoms other than oxygen must be coupled to the oxygen octahedra tilts. For example, the presence of the correlated displacements of these atoms in 011 type directions can explain the appearance of the α spots. In order to double the unit cell, these displacements must be arranged in some kind of anti-parallel fashion.

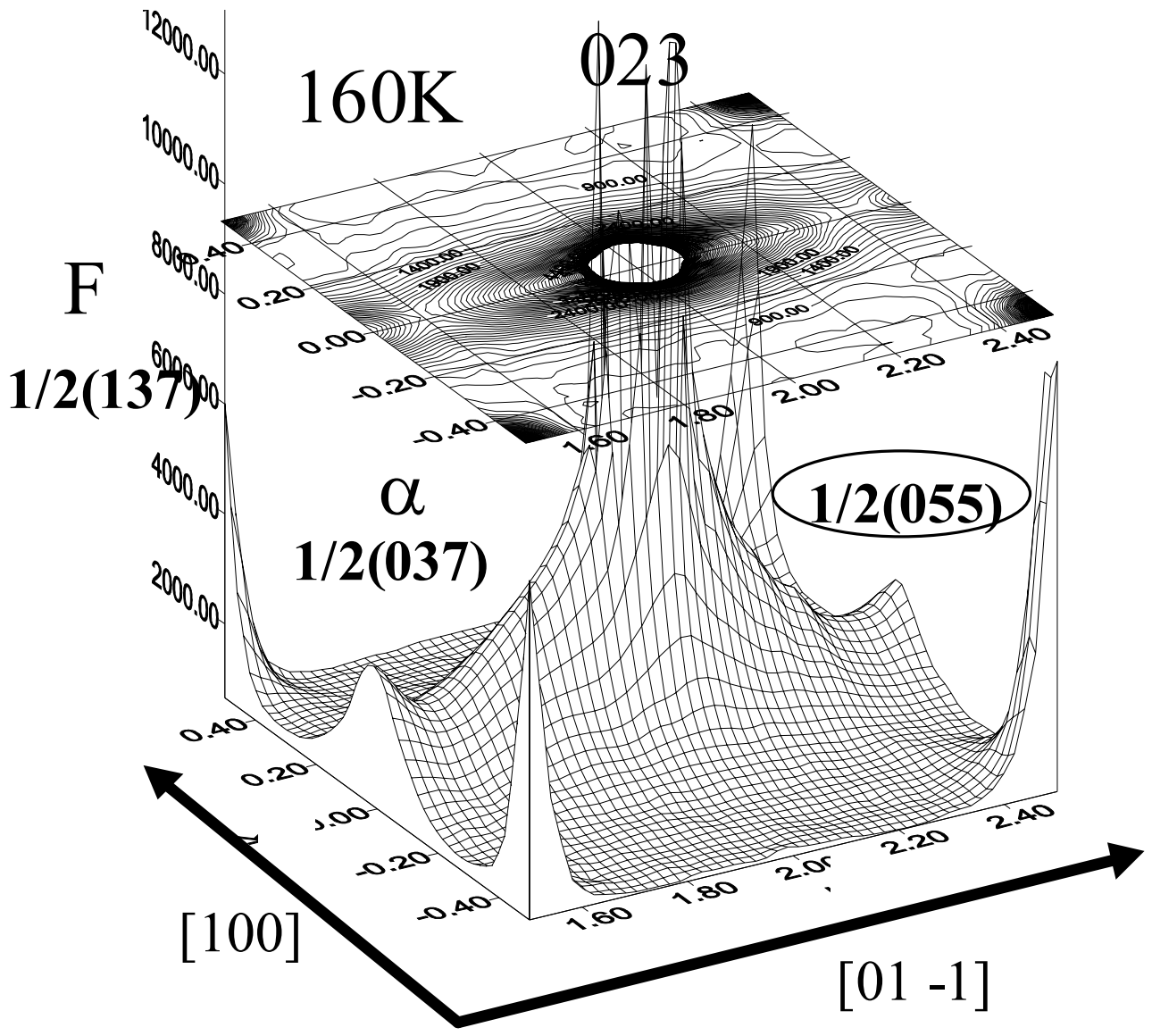

**FIGURE 5.** Superlattice peaks near the 023 Bragg paek. 1/2(055) α spot is circled. α spots with two odd Miller indices the same cannot occur due to the oxygen tilting alone [21].

These displacements are anti-ferrodistortive in nature. In a related $PbIn_{1/2}Nb_{1/2}O_3$ (PIN), α spots were also found and their temperature dependence was directly correlated with anti-ferroelectric behavior [22]. For example, the intensity of the α spots in ordered PIN sharply increased just below the Neel temperature of the anti-ferroelectric phase transition. Assuming that in PMN-PT α spots are also due to anti-ferrodistortions, we can make some interesting conclusions from the analysis of PMN and PMN-6%PT. These correlated anti-ferrodistortions form clusters whose size is temperature independent and differ from the chemical nanodomains. FWHM along the ridge (see Figure 3a) of pure PMN and PMN-6%PT is both identical and temperature independent at temperatures where α peaks can be distinguished from the diffuse background, as shown in Figure 6.

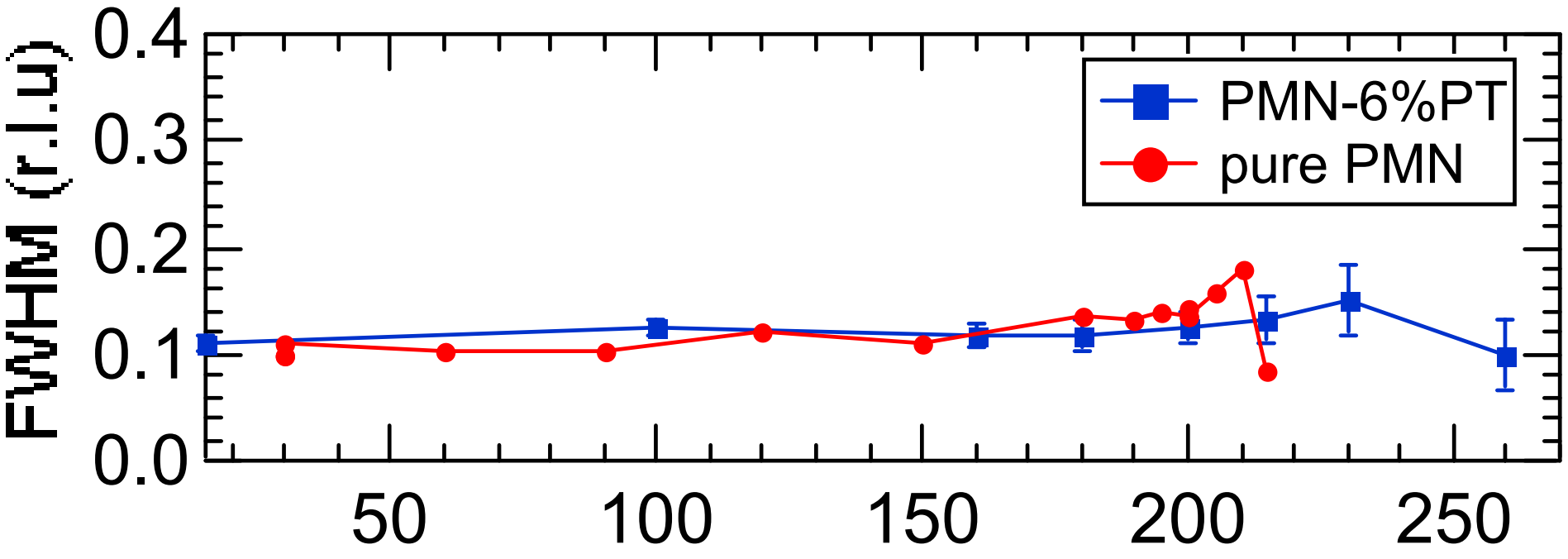

**FIGURE 6.** FWHM for 1/2(035) α spot along the ridge for PMN and PMN-6%PT.

Figure 6 suggests the size of the clusters formed by anti-ferrodistortive fluctuations is an intrinsic property independent of these two different compositions. The size was estimated to be near 32 Å for both PMN and PMN-6%PT from the Scherrer formula.

On the plot of the integrated area versus temperature (see Figure 4) the interpolated temperature, where intensity goes to zero, was found close to $T_f$=260 K in PMN-6%PT and $T_f$=220 K for pure PMN. Interestingly, these are the freezing temperatures that are very close to the ones obtained from the Vogel-Fulcher fit of the frequency dependent dielectric response. Vogel-Fulcher fits from the dielectric measurements are shown in Figure 7. From Figure 4 it is also evident that intensity of the $\alpha$ peaks in PMN-6%PT decreases more gradually than in pure PMN. Long tails are often indicative of the short-range order effects.

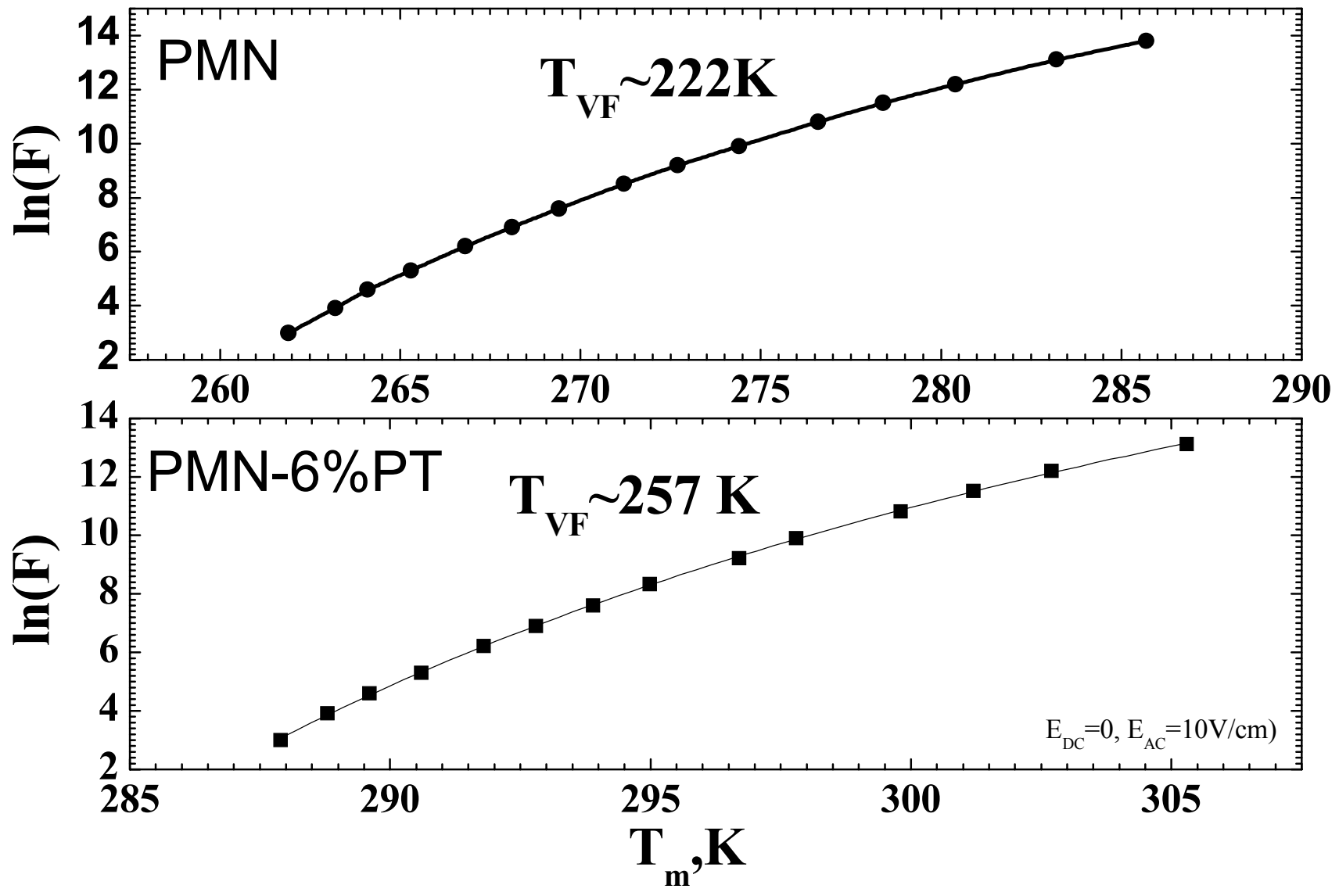

**FIGURE 7.** Plots of the frequency dependence of the dielectric constant maximum $T_m$ for PMN and PMN-6%PT. Temperature ($T_{VF}$), indicated on the plots, was used as an adjustable parameter in Vogel-Fulcher fitting equation.

# CONCLUSIONS

The temperature dependence of superlattice reflections for three compositions of PMN (with 0%, 6% and 32 % $PbTiO_3$) has been determined by utilizing synchrotron scattering techniques. No superlattice peaks of any kind were found for the 32% composition. We have found the temperature dependence of $\alpha$ spots by separating them from the temperature independent anisotropic diffuse scattering background. By doing so, we have found that their integrated intensity increases on cooling below the freezing temperature $T_f$, while FWHM of the $\alpha$ spots is temperature independent. This means that magnitude of the correlated anti-parallel atomic distortions increases on cooling below the freezing temperature, while the size of the clusters that are formed by these distortions is temperature independent. The size of these clusters is about 32 Å, and is also identical for both 0% and 6% PT doped compositions. However, the size of the chemically ordered nanodomains is twice as large for pure PMN compared to PMN-6%PT (50 Å and 24 Å respectively). This suggests that chemical ordering is independent of the clusters that are responsible for $\alpha$ spots. The temperature $T_f$ obtained from our x-ray studies is close to the phenomenological $T_f$ freezing temperature obtained from the Vogel-Fulcher fit of the frequency dependence of the $T_m$ maximum of the dielectric constant. It is not clear what the connection is between the anti-polar clusters and polar nanodomains expected in relaxor ferroelectrics, because the size of anti-ferrodistortive clusters we found to be temperature independent. On the other hand, the polar nanodomains are expected to grow in size from embryo size at $T_d$=600 K to the cryogenic temperatures, reaching couple of hundreds of Å in size. We can only speculate that these anti-polar displacements correlated on a 30 Å scale, may compete with establishment of the parallel alignment of electric dipoles which is required to catalyze a ferroelectric phase. Further studies are underway in effort to determine which atoms, apart from the oxygen, contribute to the structure factor of the $\alpha$ spots from the anomalous scattering synchrotron techniques.

## ACKNOWLEDGMENTS

This material is based upon work supported by the U.S. Department of Energy, Division of Materials Sciences under Award No. DEFG02-ER9645439, through the Frederick Seitz Materials Research Laboratory at the University of Illinois at Urbana-Champaign.

We would like to thank the staff of both the UNICAT ID-33 beamline and the MATRIX X-18A beamline at National Synchrotron Light Source, Brookhaven National Laboratory for assistance during the experiments.

The UNICAT facility at the Advanced Photon Source (APS) is supported by the Univ of Illinois at Urbana-Champaign, Materials Research Laboratory (U.S. DOE, the State of Illinois-IBHE-HECA, and the NSF), the Oak Ridge National Laboratory (U.S. DOE under contract with UT-Battelle LLC), the National Institute of Standards and Technology (U.S. Department of Commerce) and UOP LLC. The APS is supported by the U.S. DOE, Basic Energy Sciences, Office of Science under contract No. W-31-109-ENG-38.

Research carried out (in part) at the National Synchrotron Light Source, Brookhaven National Laboratory, which is supported by the U.S. Department of Energy, Division of Materials Sciences and Division of Chemical Sciences.